\documentclass{article}

\begin{document}

\title{Open Problems in Nuclear Density Functional Theory}

\author{B. G. Giraud \\
bertrand.giraud@cea.fr, Institut de Physique Th\'eorique, \\
Centre d'Etudes Saclay, 91190 Gif-sur-Yvette, France}

\date{\today} 
\maketitle

\begin{abstract}
This note describes five subjects of some interest for the density functional
theory in nuclear physics. These are, respectively, i) the need for concave
functionals, ii) the nature of the Kohn-Sham potential for the radial density
theory, iii) a proper implementation of a density functional for an 
``intrinsic'' rotational density, iv) the possible existence of a potential
driving the square root of the density, and v) the existence of many
models where a density functional can be explicitly constructed.
\end{abstract}

\bigskip
{\it Preliminary considerations} : The nuclear Hamiltonian used here reads,
\begin{equation}
H=\sum_i \frac{p_i^2}{2m}+\sum_{i>j} v_{ij}+ \varepsilon 
\left(\sum_i \vec r_i\right)^2+\alpha\, {\bf N}^2+2\beta\, {\bf N}\, {\bf Z}
+\gamma\, {\bf N}^2,
\label{Hamil}
\end{equation}
where $\vec r_i$ and $\vec p_i$ are the coordinates and momenta of identical 
nucleons, with spin and isospin labels understood unless necessary, $m$ is 
their mass, $v_{ij}$ is the two-body interaction, Galilean invariant. A term,
$\sum_{i>j>k} w_{ijk},$ for three-body interactions, can be added if needed, 
provided it is also Galilean invariant. The role of the additional terms, a 
center-of mass trap with strength $\varepsilon,$ and particle number 
fluctuation terms, with a positive definite matrix, 
$\left[\matrix{\alpha & \beta \cr \beta & \gamma}\right],$ will be clarified 
in time in the following. The operators ${\bf N}$ and ${\bf Z}$ are the
neutron and proton number operators. Since they commute with the physical
part of $H,$ namely $H_{phys}=\sum p_i^2+\sum v_{ij},$ their presence leaves
the eigenstates intact and only modifies the spectrum by trivial quantities,
which can be easily subtracted eventually. Similarly, the center-of-mass trap
commutes with that internal dynamics governed by the internal Hamiltonian, 
$H_{int}=\sum p_i^2/(2m)-\left(\sum \vec p_i\right)^2/(2 A m)+\sum v_{ij},$ 
where $A \equiv N+Z$ is the mass number. The internal dynamics is not 
perturbed, the center-of-mass motion factorizes out and the corresponding
zero-point energy can be trivially subtracted eventually.

\section{Need for concavity}

At the core of density functional (DF) theory, there is the process, analyzed
by Levy and by Lieb \cite{LevLie}, of a minimization of energy under the
constraint of a given density,
\begin{equation}
E(N,Z)={\rm Min}_{\rho=>N,Z}\ F[\rho],\ \ \ 
F[\rho]={\rm Min}_{{\cal D}=>\rho}\ {\rm Tr}\, H\, {\cal D}.
\label{definDF}
\end{equation}
Here ${\cal D}$ is any physical density operator in the Fock many-body space,
the nuclear Hamiltonian $H$ is second quantized if necessary, the trace 
${\rm Tr}$ is taken in the many-body space, the arrows $=>$ are shorthand for 
the constraints,
\begin{equation}
\int d\vec r\, \rho_n(\vec r)=N,\ \int d\vec r\, \rho_p(\vec r)=Z,\ \ \ \
\rho(\vec r)={\rm Tr}\, a^{\dagger}_{\vec r}\, a_{\vec r}\ {\cal D}.
\label{cnstrnt}
\end{equation}
where $a^{\dagger}_{\vec r}$ and $a_{\vec r}$ are the usual creation and
annihilation operators at position $\vec r,$ with spin and isospin labels 
understood. It is also understood that $\rho$ has two components, a neutron 
and a proton one. Spins are most often summed upon. The minimization, 
${\rm Min},$ may be understood as an infimum, ${\rm Inf},$ if necessary,
depending on fine details of the considered density operators, projectors 
${\cal D}=| \Psi \rangle \langle \Psi |$ or, in obvious notations, ensemble
mixtures ${\cal D}=\sum_n w_n\, | \Psi_n \rangle \langle \Psi_n |.$ Fine
details of the representability of $\rho$ are also understood in this paper.

It is well known, from the Hohenberg-Kohn formulation \cite{HK}, that the 
constraint, ${\cal D}=>\rho,$ can be implemented by means of auxiliary 
potentials $u_n,u_p,$ with Lagrange terms, 
$\int d \vec r\, u_n(\vec r)\, \rho_n(\vec r),$
$\int d \vec r\, u_p(\vec r)\, \rho_p(\vec r),$ and subsequent Legendre 
transforms.
 
Before the true spectrum of $H$ exhibits simultaneous eigenstates of ${\bf N},$
${\bf Z}$ and $H_{int},$ nothing prevents the theory from considering density
operators ${\cal D}$ that are not eigenstates of the particle number operators.
It is therefore also useful to consider, for instance, pairing densities for
opposite spin particles, 
$\kappa(\vec r)=
{\rm Tr}\, a^{\dagger}_{\vec r\, +}\, a^{\dagger}_{\vec r\, -}\, {\cal D},$ 
and minimize the energy under both the $\rho$ and the $\kappa$ constraints. 
This generalizes functionals $F[\rho]$ into functionals $F[\rho,\kappa].$ See
\cite{Bulg} for specificities associated with zero range pairing.

A mass formula is concave if any second difference such as, 
$E(N+1,Z)-2\, E(N,Z)+E(N-1,Z),$ $E(N,Z+1)-2\, E(N,Z)+E(N,Z-1),$ and similarly
with any range in any direction of the $N,Z$ plane, is positive definite. It
is a fact of life, however, that the table of nuclear ground state energies is
far from providing systematically positive second differences. Not only there 
is a significant staggering effect in binding when comparing neighboring
odd and even nuclei, because of pairing, but even more severe deviations of
concavity occur almost every time a magic shell or subshell closes.

This lack of concavity has a severe consequence for a mass formula such as
that provided by Eq. (\ref{definDF}). Assume indeed that, in the neighborhood
of an $N,Z$ nucleus, there are two integers ${n,z}$ so that the experimental 
second difference of ground state energies, 
$E_{N+n,Z+z}-2\, E_{N,Z}+E_{N-n,Z-z}$ is negative. Let $\Psi_{N+n,Z+z}$ and
$\Psi_{N-n,Z-z}$ be corresponding ground state wave functions. (If ground
states degenerate because of a spin different from zero, any magnetic label
will do.) Then the ensemble mixture, 
$${\cal D}=\frac{1}{2} \left(
| \Psi_{N+n,Z+z} \rangle \langle \Psi_{N+n,Z+z} | + 
| \Psi_{N-n,Z-z} \rangle \langle \Psi_{N-n,Z-z} | \right),$$
induces a density, $\frac{1}{2} \left(\rho_{N+n,Z+z}+\rho_{N-n,Z-z}\right),$
with correct average particle numbers $N$ and $Z.$ But it returns an absurd 
energy, $\frac{1}{2}\left(E_{N+n,Z+z}+E_{N-n,Z-z}\right),$ lower than
$E_{N,Z}.$

In Eq. (\ref{Hamil}), the terms quadratic in ${\bf N}$ and ${\bf Z},$ however,
do create concavity, provided the eigenvalues of 
$\left[\matrix{\alpha & \beta \cr \beta & \gamma}\right]$ are large enough.
A rough inspection \cite {BGJT} of the table of nuclear bindings indicates
that eigenvalues of a few MeV at most are enough to ensure concavity.

One might argue that the Kohn-Sham (KS) determinant \cite{KS} defines
particle numbers well, but the argument does not extend to DF calculations
with pairing. Indeed, procedures for particle number ``projection'' had to
be designed \cite{Doba1} \cite{Bend}. It might be that energy density
functional (EDF) theory \cite{EDF}, with its several quasi local densities,
is protected against such a syndrome related to ensemble densities, because
EDFT carries more information about the nucleus, but the question of
concavity, up to my knowledge, has not been discussed explicitly. Nor has it
been raised in the ``no Hamiltonian'' context. 

It is clear that the ``quadratic {\bf N}, ${\bf Z}$ terms'' in 
Eq. (\ref{Hamil}) are there to penalize particle number fluctuations. They 
relocate energy minima at strict eigenvalues of ${\bf N}$ and ${\bf Z}.$ This
seems to make an argument to advocate Hamiltonian approaches to DFT and be
cautious about non-Hamiltonian ones, incidentally. Anyhow a problem remains :
how does one implement such terms in the construction of a DF? It is likely
that a formalism including pairing correlations $\kappa,$ sensitive to the
presence of such terms, will be needed. The precise adjustment 
of DF and EDF theories to rigorously ensure concavity is an open problem.

\section{Radial Kohn-Sham theory}

Whether one considers the usual Hamiltonian $H_{phys},$ or the internal one, 
$H_{int},$ or that full $H,$ see Eq. (\ref {Hamil}), which is adapted to DFT, 
such operators are scalars under rotation. Accordingly, the operation, 
${\rm Tr}\, H\, {\cal D},$ which defines the energy, see
Eq. (\ref{definDF}), is sensitive to only the scalar part of ${\cal D}.$ 
(Notice that the normalization, ${\rm Tr}\, {\cal D}=1,$ also selects the 
scalar part of ${\cal D}.$) The energy minimization can, therefore, be 
restricted to a subset of purely scalar ${\cal D}$s, without any loss of 
information \cite{Girscal}. For nuclear ground states that have a non zero
spin $J,$ the degenerate magnetic multiplet can be combined into a scalar
density operator, the mixture 
${\cal D}_0=\sum_M |JM \rangle \langle JM |/(2J+1), $ with, obviously, the
ground state energy given by, $E_J={\rm Tr}\, H_{int}\, {\cal D}_0.$
The same mixture results from DFT at finite temperature \cite{Mermin}, at the
zero temperature limit.

Since the energy minimization occurs within a subspace of scalar density 
operators, the corresponding one-body densities $\rho(\vec r)$ are just
radial profiles, $\rho_0(r).$ The final result is a scalar functional,
$F_0[\rho_0].$ Hence an open problem : for such monopolar profiles $\rho_0,$
what is the signature for deformation? It is likely that the slope of 
$\rho_0$ at the nuclear surface will be weaker for deformed nuclei than for
spherical ones, but precise criteria distinguishing hard from soft nuclei are
desirable, not to mention criteria identifying triaxial nuclei, halo nuclei,
etc. Can one use properties of moments of the profiles? Such a theory with 
radial profiles, in one dimension rather than the three dimensions of usual
nuclear (E)DF theories, provides a considerable simplification of numerics,
but demands somewhat subtle criteria for the identification of deformations.

Another open problem consists in adapting the KS procedure to this radial
world. Slater determinants, made of LS orbitals, $| nlm \sigma \rangle,$ or
jj ones, $| nljm \rangle,$ orbitals provided by scalar mean fields $u_0(r),$
are never scalar wave functions, except at subshell closure. It seems
therefore necessary to redefine the KS kinetic DF from a formula similar to
Eq. (\ref{definDF}),
\begin{equation}
F_{KS0}[\rho_0]={\rm Min}_{{\cal D}_0'=>\rho_0}\ 
{\rm Tr}\, \left[\sum_i p_i^2/(2m)\right] {\cal D}_0',
\label{redefinKS}
\end{equation}
where ${\cal D}_0'$ means scalar mixtures of determinants made of the LS 
or jj orbitals. For instance, given a neutron number $N,$ assume a filled 
neutron core accommodating $N'$ neutrons and an open subshell with spin $j$
to accommodate the $(N-N')$ neutrons left. Then ${\cal D}_0'$ can represent
a mixture, with equal weights, of all combinations of $(N-N')$ orbitals within
the $(2j+1)$ available ones. For a justification of the equal filling 
approximation, see \cite{GOKPMR}.

Then a KS potential $u_{KS0}$ can be defined as,
\begin{equation}
r\, u_{KS0}(r)=\frac{\delta (F_0-F_{KS0})}{r\, \delta \rho_0(r)},
\end{equation}
and can be further simplified if a Hartree subtraction makes sense for nuclear
forces. Finally a spherical set of orbitals must be generated, from the 
potential $u_{KS0},$ to obtain self-consistency of $\rho_0.$

Little is known at present in nuclear physics about this redefined kinetic 
DF, Eq. (\ref{redefinKS}). It is time to import, and maybe readjust, the 
results that are known in atomic physics \cite{Nagy} \cite{UK}. While the 
early literature of nuclear physics is not devoid of effective spherical
potentials for shell models, and while there has been recently interesting
attempts to describe spherical nuclei with empirical EDFs, see for instance
\cite{Schuck1}, very little is known about $u_{KS0}$  either. It
may be also worth comparing such ``spherical'' EDFs with those used
for deformed nuclei, see for instance \cite{Schuck2}. The program of a radial
KS theory is a largely open problem.

\section{Functionals for intrinsic rotational states}

Let us first avoid a frequent semantic confusion between ``intrinsic'' and 
``internal''. In this note, the word ``intrinsic'' shall be used for a 
state $\Phi,$ which has to be {\it simple}, and out of which rotational
states can be obtained. Simplicity means, in particular, that one uses 
microscopic degrees of freedom, in the laboratory frame. The state $\Phi$ is
typically a Slater determinant or a Hartree-Bogoliubov state. Rotational
states are calculated from the usual algebra with rotation operators and
rotation matrices. Skipping all technical details, the main formula in this
algebra reads, in transparent notations,
\begin{equation}
\Phi_{MK}^J = \int d\psi d\theta d\varphi\, D^J_{MK}(\psi,\theta,\varphi)\,
\exp (-i \psi {\bf J}_z) \exp(-i \theta {\bf J}_y) \exp(-i \varphi {\bf J}_z)\,
| \Phi \rangle.
\end{equation}
(If necessary, a subsequent configuration mixing, 
$\Psi=\sum_K\, c_K\, \Phi^J_{MK},$ may improve the calculation of a physical
state.) The main point of interest is that the states $\Phi^J_{MK},$
because of their quantum numbers, contain correlations that $\Phi$ {\it must
not} contain in order to be ``simple''.

It is well known that, except at the limit of very hard rotators, there is no
decoupling possible between the three rotation degrees of freedom and any
residual set of $(3 A-3)$ ``internal'' degrees of freedom describing any
``internal'' motion inside the rotator. Such is not the case for the 
translation group, where the internal Hamiltonian,
$H_{int}=\sum p_i^2/(2m)-\left(\sum \vec p_i\right)^2/(2 A m)+\sum v_{ij},$
and the trapped center-of-mass one, 
$H_{com}=\left(
\sum \vec p_i\right)^2/(2 A m)+\varepsilon\, \left(\sum \vec r_i\right)^2,$
drive separate dynamics. The subtle relation between internal density and
density in the laboratory and the theories for the corresponding DFs make a
closed problem  \cite{com}, which is, therefore, out of the
scope of this note. The word ``internal'' shall be kept for the context of the
translation group and the word ``intrinsic'' shall be used here, specifically,
in the context of the rotational group.

Assume that every $\Phi^J_{MK}$ can be a good approximation to an eigenstate
of $H$ and consider the two-step minimization,
\begin{equation}
E_{JMK}={\rm Min}_{\rho}\, F_{JMK}[\rho],\ \ \ \ 
F_{JMK}[\rho]={\rm Min}_{\Phi=>\rho}\,
\frac{\langle \Phi^J_{MK} | H | \Phi^J_{MK} \rangle}{\langle \Phi^J_{MK} | 
\Phi^J_{MK} \rangle},
\end{equation}
where $\rho$ is the density of $\Phi,$ and $\Phi$ is {\it restricted} to
a class of simple states. This defines a density functional theory
\cite{Girintrnsc} in terms of the intrinsic density, not the density of that 
(hopefully physical) state $\Phi^J_{MK}$ which provides the energy. But, 
obviously, this DF, $F_{JMK}[\rho],$ depends on that subset to which $\Phi$ is
restricted. Moreover, it depends on $J,M,K.$

There are not many subsets available for $\Phi.$ Hartree-Bogoliubov solutions
do not make a bad first choice. But then the essentially desirable virtue of 
$\Phi$ would be to be practically independent of $J,M,K.$ Can this be true?
Hence the open problem : under which conditions can $F_{JMK}$ define just
one density $\rho$ valid for a large number of combinations $JMK$? How does
this intrinsic DF relate \cite{Doba2} to the traditional three-dimensional
(E)DFs used for deformed nuclei?

\section{About the square root of the density}

A theorem, studied by many authors in atomic or molecular physics, see in
particular \cite{LevPerSah}, states that, given the density $\rho(\vec r)$
of an eigenstate of $H$ with energy $E_A,$ there exists a local potential,
$u_{sqr},$ that drives the square root of $\rho,$
\begin{equation}
[-\hbar^2 \Delta_{\vec r}/(2 m)+u_{sqr}(\vec r)]\, \sqrt{\rho(\vec r)}=
(E_A-E_{A-1})\, \sqrt{\rho(\vec r)}\, .
\end{equation}
Here $E_{A-1}$ is the ground state energy of the system obtained
by removing one particle. (When $E_A$ corresponds to the ground state, the
difference, $(E_A-E_{A-1}),$ is a separation energy.) The details of the
proof, found in \cite{LevPerSah}, will not be discussed here. The theorem was
recently generalized for mixtures of degenerate states \cite{GirMou}. Let us
only mention that part of $u_{sqr}$ consists in a Hartree-like convolution of
the two-body interaction $v_{ij}$  with a density in the $(A-1)$ space.
The theorem was established for electrons in atoms or molecules, where
$v_{ij}$ is the Coulomb repulsion, local. Locality plays a role in the 
proof of the theorem. It seems, therefore, that an extension of the theorem
with nuclear interactions, which can be seriously non local, is excluded.
But the nuclear DFT makes an extensive use of locality or quasi-locality.
The KS potential, in particular, is a functional derivative, local. Moreover,
effective forces can be local, optical potentials can be local, Hartree
potentials can be local, up to reasonable approximations at least. Can there
be a generic form of an approximate, local $u_{sqr},$ valid for many nuclei?
Would it relate to the nuclear DFT?

\section{Constructive models and their polynomial zoo}

Everyone knows that existence theorems for DFs do not provide simple, explicit
constructions \cite{DG} \cite{DruFurPla}. However, a direct, constructive
approach has recently been proposed \cite{algbr}, with a toy model explaining
the construction. But this is just a toy model. For a more realistic, but
still not too unwieldy model of an algebraic implementation of the constraint, 
${\cal D}_0=>\rho_0,$ let ${\cal D}_0$ be just a projector of rank one, 
$|\Psi \rangle \langle \Psi |.$ Use the spherical harmonic oscillator with
its standard set of orbitals,
$\varphi_{nlm \sigma \tau}(\vec r)=Y_{lm}(\hat r)\, \exp(-\nu r^2/2)\, 
P_{nl}(r)\, \chi_{\sigma}\, \chi_{\tau},$ where $\sigma=\pm$ and $\tau=p,n$
are spin and isospin labels, and, more important, the orbitals have the 
same exponential decay, modulated by polynomials $P_{nl}.$ {\it Truncate
this single particle basis at some maximum value of the number of
$\hbar \omega$ quanta}. The truncation may differ for neutrons and protons,
if necessary. Given a neutron and a proton numbers, $N,Z,$ prepare an
orthonormalized shell model basis of states $\Phi_i$ for a configuration 
mixing, $\Psi=\sum_j (c_j+i\, c'_j)\, \Phi_j,$ to describe this $N,Z$ nucleus,
assumed to have a ground state spin zero. (Here the real and imaginary
parts, $c_j,c'_j$ of the mixing coefficients have been explicited.) The
$\Phi_i$'s, naturally, are made of
$N+Z$ orbitals $\varphi_{nlm\sigma\tau},$ with the necessary recoupling of
angular momenta to induce total angular momentum $0.$ Given $H,$ tabulate
the matrix elements, $H_{ij} \equiv \langle \Phi_i | H | \Phi_j \rangle,$
making most often a sparse matrix. The energy is,
\begin{equation}
\eta = \sum_{ij}\, (c_i-i\, c'_i)\, H_{ij}\, (c_j+i\, c'_j).
\label{enrj}
\end{equation}

Then the neutron density, 

\begin{equation}
\rho_n(\vec r) = \sum_{ij}\, (c_i-i\, c'_i)\, 
\langle \Phi_i |\, (a^{\dagger}_{\vec r\, +n}\ a_{\vec r\, +n} +
                    a^{\dagger}_{\vec r\, -n}\ a_{\vec r\, -n})\, | \Phi_j 
\rangle\, (c_j+i\, c'_j),
\label{neutdns}
\end{equation}
is also a polynomial in terms of the real and imaginary parts $c$'s, $c'$'s.
It is also a radial function, $\rho(r).$ Moreover, it is a polynomial of $r,$
multiplied by $\exp(-\nu r^2).$ Its order ${\cal N}_n$ is indeed finite,
since the single particle basis was truncated at some harmonic oscillator
major shell. Only ${\cal N}_n$ of the ${\cal N}_n+1$ coefficients are
independent, because the integral, $\int_0^{\infty} r^2\, dr\, \rho_n(r),$
equates the neutron number. 

Let ${\bf C}^n_m$ denote the independent coefficient multiplying $r^m$ in this
``neutron density polynomial''. It can be stressed that the matrix element, 
$\langle \Phi_i | a^{\dagger}_{\vec r\, n}\, a_{\vec r\, n} | \Phi_j \rangle,$ 
(spin summation understood) is also the product of $\exp(-\nu r^2)$ by a
polynomial of $r.$ Hence, let such matrix elements be tabulated, and,
in particular, let their coefficients $C^n_{ijm}$ of $r^m$ be tabulated. 
The ``neutron density constraint'', in the Levy and Lieb sense \cite{LevLie},
is thus the set of algebraic relations,
\begin{equation}
\sum_{ij}(c_i-i\, c'_i)\, (c_j+i\, c'_j)\ C^n_{ijm} = {\bf C}^n_m\, ,\ \ \ 
m=1,...,{\cal N}_n\, .
\label{neutcnstr}
\end{equation}

The same considerations hold for the proton density constraint, leading to 
the set of polynomial constraints,
\begin{equation}
\sum_{ij}(c_i-i\, c'_i)\, (c_j+i\, c'_j)\ C^p_{ijm} = {\bf C}^p_m\, ,\ \ \ 
m=1,...,{\cal N}_p\, .
\label{protcnstr}
\end{equation}

The last constraint to consider is the normalization of $\Psi,$
\begin{equation}
\sum_i c_i^2+(c'_i)^2=1,
\label{normcnstr}
\end{equation}
hence, the total number of constraints is, a priori,
${\cal N}_c={\cal N}_n+{\cal N}_p+1.$

A priori also, the total number of real parameters $c_i, c'_i$ is twice the
number of states $\Phi_i,$ diminished by $1,$ because of the arbitrary phase of
$\Psi.$ Accordingly, one of the $c'$'s, for instance, can be frozen to zero.
Actually, except for clumsy choices of phases in the basis, or special
Hamiltonians breaking time reversal symmetry, it is reasonable to assume
 that all $c'$'s vanish. The model can be run with $c$'s only, the number 
${\bf N}$ of which is the dimension of the basis $\{\Phi_i\}.$

In every realistic model one can think of, the number ${\bf N}$ of parameters
is significantly larger than that, ${\cal N}_c,$ of constraints. Then, because
of Eqs. (\ref{neutcnstr}), (\ref{protcnstr}) and (\ref{normcnstr}), one can
eliminate ${\cal N}_c$ among the coefficients $c_i$ from Eq. (\ref{enrj}).
This leaves a ``precursor'' polynomial relation, 
\begin{equation}
{\cal P}_{prec}(\eta,{\bf C}^n_1,{\bf C}^n_2,...,{\bf C}^n_{{\cal N}_n},
{\bf C}^p_1,{\bf C}^p_2,...,{\bf C}^p_{{\cal N}_p},c_i^{surv})=0,
\end{equation}
between the energy $\eta,$ the coefficients ${\bf C}^{n,p}_m$ of the
density polynomials and the $({\bf N}-{\cal N}_c)$ ``survivor'' coefficients
$c_i^{surv}.$ Then the minimization of $\eta$ under the constraint of a given
density \cite{LevLie} induces $({\bf N}-{\cal N}_c)$ equations,
$\partial {\cal P}_{prec}/\partial c_i^{surv}=0.$ These are used in turn to
eliminate the survivors. Every step in the process is an easy manipulation
of polynomials. One finally obtains a polynomial relation,
\begin{equation}
{\cal R}(\eta,{\bf C}^n_1,{\bf C}^n_2,...,{\bf C}^n_{{\cal N}_n},{\bf C}^p_1,
{\bf C}^p_2,...,{\bf C}^p_{{\cal N}_p})=0,
\end{equation}
which links the minimal energy to the parameters of $\rho.$ Although not an
open formula, this implicit equation, ${\cal R}=0,$ has all the needed virtues
of a density functional for the calculation of $\eta$ and its minimization in 
terms of the density. 

This detour via the density to avoid a direct diagonalization of the matrix
$\{H_{ij}\}$ is of some interest if the matrix is sparse but too huge to be
handled directly. It is not excluded, in particular, that those polynomials
${\cal R}$ obtained from ``small'' models turn out to give reasonable
approximations for ``bigger'' models. It is indeed likely that physics allows
a natural cut-off in the order of the density polynomials, because high
orders might mean oscillations that are too costly in energy to be acceptable
for ground states.

It is clear that this rigorous derivation raises several technical problems,
such as, for instance, the best tactics to handle big polynomials in the
context of many parameters, the comparison of distinct polynomials ${\cal R}$
obtained from models with different choices for the basis, the search for
extrapolations when calculations become too cumbersome. But there is no doubt
that this constructive approach offers challenging open problems.

An interesting side question arises: can this algebraic method also provide
the KS potential? We have preliminary results which are optimistic, but they
are only ... preliminary. Another interesting question is, can this algebraic
method be used for the cluster model? We have again preliminary results which
are optimistic. But they are also ... too preliminary.

Last, but not least, it can be noticed that this polynomial method can be used
for other collective degrees of freedom than the density. Every quadrupole, 
octupole, etc, indeed boils down to a polynomial with respect to the configuration
mixing coefficients.

\section{ Discussion and Conclusion} 
Five problems have been stated by this note. The most urgent seems to be
the concavity question. Constrained minimizations in a landscape that is
not concave can lead to absurd mistakes.

Then comes the question of the radial theory. The empirical success of the
present three-dimensional approaches, EDF in particular, should not hide the
fact that these contradict the rotational invariance of the Hamiltonian. 
The excuse that mean field theory can break symmetry is correct, but is not
acceptable for a rigorous DFT. Density functional theory is ${\it not}$ a
mean field theory, despite the Kohn-Sham formulation. This KS formulation
must be readapted to the scalar world demanded by the scalar nature of $H.$

It might be that three-dimensional approaches can be justified within the 
third subject discussed by this note, a variational principle based upon
variation after projection of good quantum numbers. But this is yet an 
open subject.

Unless we learn of further evidence, the fourth subject, about the square root
of the density, does not seem to deserve priority. Still, it would relate
various effective potentials, of which nuclear physics is an active consumer.

The last subject, a constructive method via polynomials, stands out as 
special. It uses a highly non local parametrization of the density, that
deviates from the (quasi) local tradition of the field. This untraditional
parametrization of $\rho,$ in the frame of a polynomial algebra, introduces
a completely new zoology of polynomial DFs. Everything must be reinvented
in this context, for the number of soluble models is huge, calculations can
be heavy and it is not obvious how to take advantage of such rigorous
solutions ... that might be sometimes obscure. But isn't this a stimulating
situation?

\medskip
Acknowledgements : It is a pleasure to thank the organizers of this exercise 
``Open Problems'' for an invitation to write this note. It is also a pleasure
to acknowledge collaborations with B.R. Barrett, B.K. Jennings,
S. Karataglidis, P. Moussa and J. Svenne and stimulating discussions with
K. Bennaceur, A. Bulgac, T. Duguet, N. Michel, P. Schuck and L. Wilets.

\end{document}